\documentclass[11pt,twoside]{article}
\usepackage{./asp2010}
\usepackage{graphicx}

\resetcounters

\begin{document}

\title{Multi-Wavelength Implications of the Companion Star in $\eta$ Carinae}
\author{Thomas~I.~Madura$^1$, Theodore~R.~Gull$^2$, Jose~H.~Groh$^1$, Stanley~P.~Owocki$^3$, Atsuo~Okazaki$^4$, D.~John~Hillier$^5$, and Christopher~Russell$^3$}
\affil{$^1$Max-Planck-Institut f\"{u}r Radioastronomie,\\ 
Auf dem H\"{u}gel 69, D-53121 Bonn, Germany\\
$^2$Astrophysics Science Division, Code 667,\\
NASA Goddard Space Flight Center, Greenbelt, MD 20771, USA\\
$^3$Department of Physics and Astronomy,\\
University of Delaware, Newark, DE 19716, USA\\
$^4$Faculty of Engineering, Hokkai-Gakuen University,\\
Toyohira-ku, Sapporo 062-8605, Japan\\
$^5$Department of Astronomy, 3941 O'Hara Street,\\
University of Pittsburg, Pittsburg, PA 15260, USA}

\begin{abstract}
$\eta$ Carinae is considered to be a massive colliding wind binary system with a highly eccentric ($e \sim 0.9$), 5.54-yr orbit. However, the companion star continues to evade direct detection as the primary dwarfs its emission at most wavelengths. Using three-dimensional (3-D) SPH simulations of $\eta$ Car's colliding winds and radiative transfer codes, we are able to compute synthetic observables across multiple wavebands for comparison to the observations. The models show that the presence of a companion star has a profound influence on the observed HST/STIS UV spectrum and H-alpha line profiles, as well as the ground-based photometric monitoring. Here, we focus on the Bore Hole effect, wherein the fast wind from the hot secondary star carves a cavity in the dense primary wind, allowing increased escape of radiation from the hotter/deeper layers of the primary's extended wind photosphere. The results have important implications for interpretations of $\eta$ Car's observables at multiple wavelengths.
\end{abstract}

\section{Introduction}
The 5.54-year periodicity observed in numerous spectral lines, as well as the IR, Visual, and X-ray fluxes, strongly suggests $\eta$ Carinae is a binary system \citep{whitelock04, damineli08a, damineli08b, eduardo10, corcoran11}. Recently, \citet[][hereafter O08]{okazaki08} modeled $\eta$ Car's \emph{RXTE} X-ray light curve using a three-dimensional (3-D) Smoothed Particle Hydrodynamics (SPH) simulation of the binary wind-wind collision. A key point of O08 is that the fast wind of the secondary star, $\eta_{B}$, carves a cavity in the dense wind of the primary, $\eta_{A}$, allowing X-rays that would otherwise be absorbed to escape into our line-of-sight. This immediately suggests that if the primary wind is sufficiently optically thick in the UV, Optical, or IR waveband, the low-density secondary wind may likewise carve or ``bore" a cavity or ``hole" in the associated wind photosphere, allowing increased escape of radiation from the hotter/deeper layers. 

\section{The Bore-Hole Effect}

Any bore-hole effect should depend on (1) how close the wind cavity carved by $\eta_{B}$ gets to $\eta_{A}$ and (2) the apparent size of $\eta_{A}$'s photosphere. For a bore-hole effect to occur, the optically thick primary wind photosphere must extend \emph{at least} as far as the separation distance between the primary and the apex of the wind-wind interaction region (WWIR), i.e. if at some point (2) $>$ (1), there is a bore hole.

The separation between $\eta_{A}$ and the apex of the WWIR is determined by the dynamics of the colliding winds and the orbit. The absolute minimum distance that $\eta_{A}$'s photosphere must extend is the separation between $\eta_{A}$ and WWIR apex at periastron:

\begin{equation}
R_{\mathrm{min}} \equiv \frac{a(1 - e)}{1 + 1/\sqrt{\eta}} \ , \label{1.1}
\end{equation}\\
where $e$ is the eccentricity (0.9), $a$ is the orbital semi-major axis length ($15.4 \ \mathrm{AU}$), and $\eta$ is the momentum ratio of $\eta_{A}$'s wind to $\eta_{B}$'s wind. For the parameters of O08, $\eta \approx 4.2$ and $R_{\mathrm{min}} \approx 1 \ \mathrm{AU}$. Using the primary mass-loss rate $\dot{M} = 10^{-3} M_{\odot} \ \mathrm{yr}^{-1}$ suggested by \citet[][hereafter H01 and H06]{hillier01, hillier06}, $\eta \approx 16.67$ and $R_{\mathrm{min}} \approx 1.24 \ \mathrm{AU}$.

The apparent size of $\eta_{A}$'s wind photosphere is determined by radiative transfer effects. While there are many ways to define this apparent size, a simple way is to use the radial photospheric radius $R_{\mathrm{phot}}$ at which the radial optical depth $\tau = 1$:

\begin{equation}
R_{\mathrm{phot}} = \frac{\kappa \dot{M}}{4 \pi v} \ , \label{1.2}
\end{equation}\\
where $\kappa$ is the opacity (units of cm$^{2}$ g$^{-1}$, assumed constant for simplicity), $\dot{M}$ is the mass-loss rate of $\eta_{A}$, and $v$ is the terminal speed of $\eta_{A}$'s wind. Assuming $\kappa = 1$ cm$^{2}$ g$^{-1}$ and mass-loss rates of $\dot{M} = 2.5 \times 10^{-4} M_{\odot}$ yr$^{-1}$ and $\dot{M} = 10^{-3} M_{\odot}$ yr$^{-1}$, $R_{\mathrm{phot}} \approx 1.67$ AU and 6.71 AU, respectively. Thus, even for low values of $\kappa$ of order unity, a bore-hole effect should occur in the $\eta$~Car system (at least at periastron).

\begin{figure}[!ht]
\begin{center}
\includegraphics[scale=0.2]{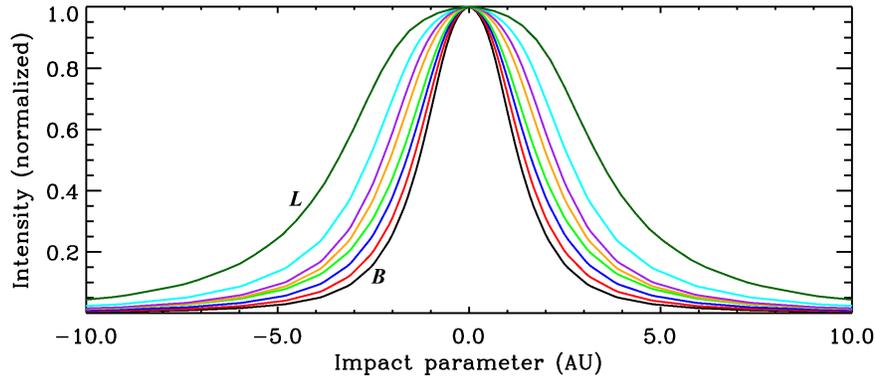}
\end{center}
\caption{Curves showing normalized intensity vs. impact parameter in each of the wavebands (from innermost to outermost) \emph{B, V, R, I, J, H, K, L}, computed using the H01, H06 \texttt{CMFGEN} model of $\eta_{A}$.}
\label{fig1}
\end{figure}

In reality, any bore-hole effect in $\eta$ Car will be wavelength dependent and occur over a relatively short period of time around periastron. The wavelength dependence is due to the opacity, which determines the size of $\eta_{A}$'s photosphere. Figure~\ref{fig1} illustrates this, showing an overlaid plot of the normalized intensity as a function of impact parameter from the center of $\eta_{A}$ in each of the wavebands \emph{B, V, R, I, J, H, K, L}. 

Since the separation between $\eta_{A}$ and the apex of the WWIR varies with orbital phase $\phi$, the bore-hole effect is also orbital-phase dependent. At apastron, the separation is $\sim 23.6 \ \mathrm{AU}$, much larger than $\eta_{A}$ in either the \emph{K}- or \emph{B}-band. Because of the large orbital eccentricity, this separation remains large for most of the orbit, $\gtrsim 10 \ \mathrm{AU}$ for $\phi$ between 0.5 and 0.925. However, between $\phi \approx 0.925$ and 1.0 (periastron), the separation drops rapidly, from $\sim 10 \ \mathrm{AU}$ to 1.24 AU. At $\phi \sim 0.97$, the separation is $\sim 5 \ \mathrm{AU}$, which is about the same size as $\eta_{A}$ in the \emph{K}-band. Therefore, for the wavebands considered here, a bore-hole effect should occur in $\eta$ Car at phases $0.97 \lesssim \phi \lesssim 1.03$.

Because of the different photospheric radii in each waveband, the wind cavity will bore different relative distances into $\eta_{A}$ at a given $\phi$ in a particular waveband. Thus, at $\phi = 0.98$, one would expect to see a significant bore-hole effect in the \emph{K}-band, but only a relatively minor effect, if any, in the \emph{B}-band. Moreover, because of the differences in relative flux probed between wavebands, one might also expect different relative changes in brightness before, during, and after periastron.

To investigate the bore-hole effect, we use 3-D SPH simulations of $\eta$ Car's colliding winds, together with a modified version of the visualization program SPLASH \citep{price07}, to generate renderings of the surface brightness of $\eta_{A}$ in each of the various wavebands \emph{B} through \emph{L} as a function of orbital phase. Opacities and primary star source functions as a function of radius in each waveband were taken from the spherically symmetric model of $\eta_{A}$ by H01, H06. Details can be found in \citet{madura10}.

\begin{figure*}[!t]\centering
  \includegraphics[width=1.0\textwidth]{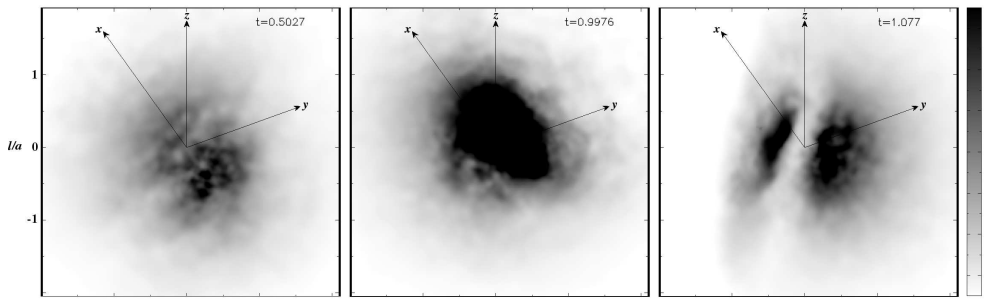}
  \caption{3-D renderings of the bore-hole effect at orbital phases of apastron (left), 5 days before periastron (middle), and 155 days after periastron (right). The  $x$ and $y$ axes are the major and minor axes, respectively, and $z$ is the orbital axis $\perp$ to the orbital plane. Lengths are in semi-major axes and grayscale indicates surface brightness. See \protect\citet{madura10} for modeling details.}
  \label{fig2}
\end{figure*}

Our models reveal that when $\eta_{B}$ is at or near apastron, the WWIR is indeed too far from $\eta_{A}$ and there is no bore hole (Fig.~\ref{fig2}, left panel). As $\eta_{B}$ moves closer to $\eta_{A}$ during its orbit, the WWIR gradually penetrates into the primary photosphere, creating a bore-hole effect that increases up until periastron passage (Fig.~\ref{fig2}, middle panel), at which point $\eta_{B}$ quickly wraps around the back side of $\eta_{A}$, the bore hole briefly vanishing as it faces away from the observer. After periastron, the bore hole reappears on the opposite side of $\eta_{A}$ (Fig.~\ref{fig2}, right panel) and then slowly fades as $\eta_{B}$ moves back toward apastron.

\begin{figure*}[!t]\centering
  \includegraphics[width=1.0\textwidth]{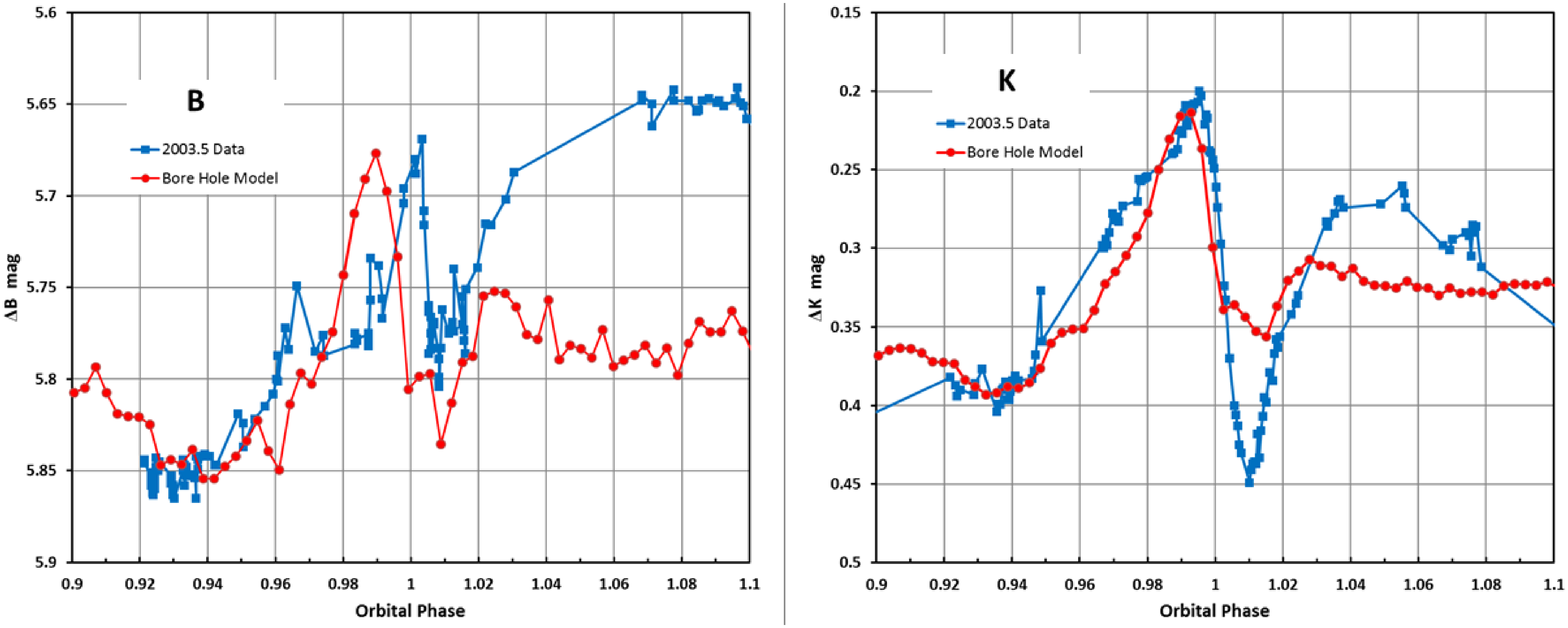}
  \caption{Synthetic light curves from the bore-hole model (circles) compared to the photometric observations of $\eta$ Car's 2003.5 event \protect\citep[squares,][]{eduardo10} in the \emph{B} (left) and \emph{K} (right) bands .}
  \label{fig3}
\end{figure*}

Using the above 3-D model, we also generate synthetic photometric light curves in each waveband for comparison to the available ground-based photometry \citep{whitelock04, eduardo10}. We find that our bore-hole model reproduces the steep rise and drop observed in each waveband before periastron, and gives roughly the same peak-to-peak change in magnitude and duration of the `eclipse-like' events seen every 5.54-years (see Figure~\ref{fig3} and Madura 2010). We thus find that $\eta_{B}$ and the WWIR have a profound influence on $\eta_{A}$'s extended photosphere and the photometric observables. The bore-hole effect likely explains $\eta$ Car's periodic eclipse-like events.

\section{Effects of a Bore Hole on \boldmath $\eta$ Car's Spectrum and Interferometric Observables}

Using the 2-D code of \citet{busche05}, we have also investigated how a bore-hole effect could alter the observed optical and UV spectra of $\eta$ Car. Details on the modeling approach can be found in \citet{groh11} and references therein.

Assuming a single-star scenario, H01 obtained a reasonable fit to the observed \emph{HST} optical spectrum obtained just after periastron, $\phi \sim 0.05$. The model spectrum from H01 reproduces well the emission line profiles of H, Fe II, and N II lines. However, compared to the observations, the H01 model overestimates the amount of P-Cygni absorption. The comparison is even worse as one moves toward apastron, when the observations show little or no P-Cygni absorption in H and Fe II lines.

Using our 2-D radiative transfer model, which takes into account the cavity in the wind of $\eta_{A}$
caused by $\eta_{B}$, we computed the synthetic optical spectrum of $\eta$ Car at $\phi \sim 0.6$ (see figure 1 of Groh 2011). We assume the same parameters for $\eta_{A}$ as H01 and a standard geometry of the WWIR at apastron as predicted by the 3-D SPH simulation. For a viewing angle with inclination $i = 41^{\circ}$ and longitude of periastron $\omega = 270^{\circ}$, the 2-D model produces a much better fit to the P-Cygni absorption line profiles of H and Fe II lines than the 1-D H01 model, while still fitting the emission line profiles. The improved fit to the P-Cygni absorption is due to the cavity in the wind of $\eta_{A}$, which reduces the H and Fe II optical depths in line-of-sight to the primary. We find similar results when modeling $\eta$ Car's observed UV spectrum.

\begin{figure*}[!t]\centering
  \includegraphics[width=0.75\textwidth]{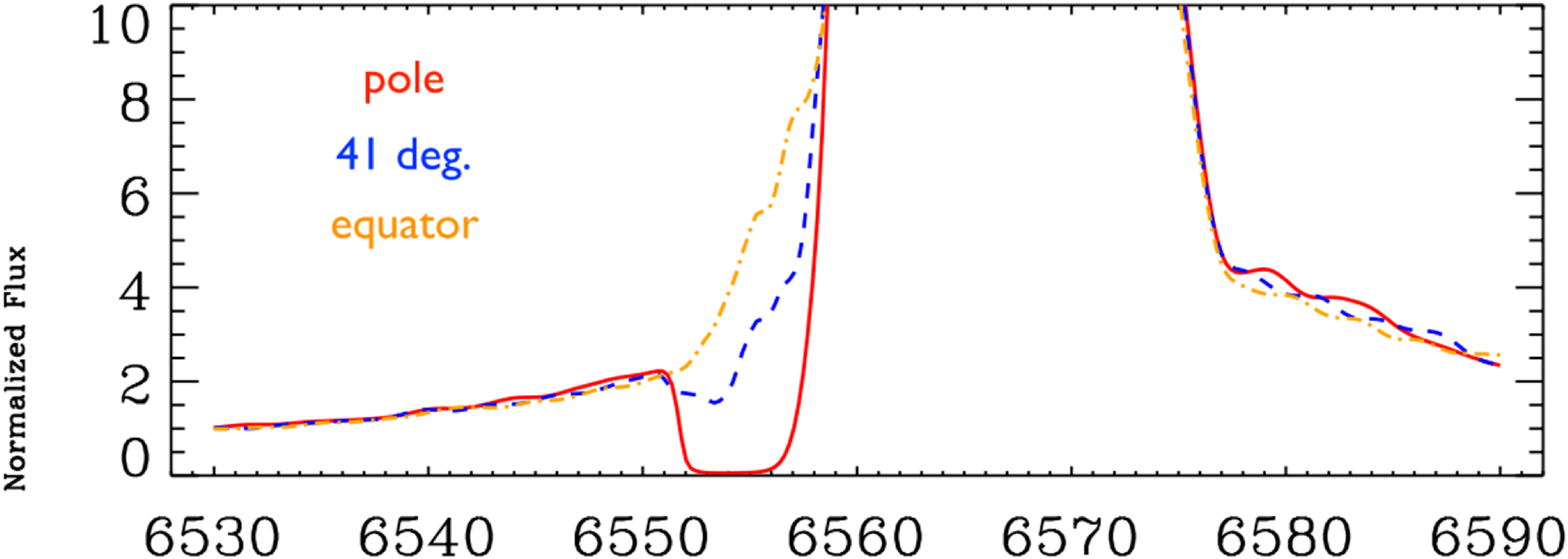}
  \caption{Synthetic line profiles of H$\alpha$ generated using the 2-D model discussed in the text, emphasizing the P-Cygni absorption at three stellar latitudes for $\eta_{A}$.}
  \label{fig4}
\end{figure*}

A preliminary study of synthetic H$\alpha$ line profiles generated using our 2-D cavity model of $\eta_{A}$ also reveals broader, deeper P-Cygni absorption near the stellar poles, compared to the equator (Figure \ref{fig4}). A bore-hole or wind-cavity model may therefore provide an alternative explanation for \emph{HST} long-slit spectral observations of H$\alpha$ line profiles as a function of stellar latitude that have been historically interpreted as evidence that the wind of $\eta_{A}$ is latitude dependent \citep{smith03}.

Finally, using a similar 2-D model, \citet{groh10} find that the density structure of $\eta_{A}$'s wind can be sufficiently disturbed by $\eta_{B}$ and the associated wind cavity, thus mimicking the effects of fast rotation in the interferometric observables. \citet{groh10} further show that even if $\eta_{A}$ is a rapid rotator, models of the interferometric data are not unique, and both prolate- and oblate-wind models can reproduce current interferometric observations. These prolate- and oblate-wind models additionally suggest that the rotation axis of $\eta_{A}$ would not be aligned with the Homunculus polar axis.

\section{Conclusions}

Our conclusions are rather straightforward; based on the results of our 3-D hydrodynamical simulations and multi-D radiative transfer modelling, we find that the secondary star $\eta_{B}$ and WWIR \emph{significantly} affect the extended wind photosphere of $\eta_{A}$, and thus multi-wavelength observations of the $\eta$ Car system.

\acknowledgements We thank Nathan Smith for discussions that led to this work.

\end{document}